# The 11-years solar cycle as the manifestation of the dark Universe

K. Zioutas [1,2], M. Tsagri [3#], Y.K. Semertzidis [4], T. Papaevangelou [5],
D.H.H. Hoffmann [6], V. Anastassopoulos [1]

[1] University of Patras, 26504 Patras, Greece.

[2] CERN, CH 1211 Geneva 23, Switzerland.

[3] Nikhef – University of Amsterdam, Amsterdam, The Netherlands.

[4] Brookhaven National Lab, Physics Deptment, Upton, NY 11973-5000, USA.

[5] IRFU, Centre d'Études Nuclaires de Saclay, Gif-sur-Yvette, France.

[6] Institut für Kernphysik, TU-Darmstadt, Schlossgartenstr. 9, 64289 Darmstadt, Germany.

**Abstract:**

Sun's luminosity in the visible changes at the $10^{-3}$ level, following an 11 years period. In X-rays, which should not be there, the amplitude varies even ~$10^5$ times stronger, making their mysterious origin since the discovery in 1938 even more puzzling, and inspiring. We suggest that the multifaceted mysterious solar cycle is due to some kind of dark matter streams hitting the Sun. Planetary gravitational lensing enhances (occasionally) slow moving flows of dark constituents towards the Sun, giving rise to the periodic behaviour. Jupiter provides the driving oscillatory force, though its 11.8 years orbital period appears slightly decreased, just as 11 years, if the lensing impact of other planets is included. Then, the 11 years solar clock may help to decipher (overlooked) signatures from the dark sector in laboratory experiments or observations in space.

[#]) present address: University of Geneva, Switzerland

Emails: hoffmann@physik.tu-darmstadt.de; thomas.papaevangelou@cea.fr; Yannis@bnl.gov; zioutas@cern.ch

1. **Introduction**

   The nearby Sun is full of large and small mysteries, with its unnatural hot outer atmosphere being the mostly impressive one, with an anomalous strong temperature rise being quasi step-like. The biggest of all mysteries, which is almost ubiquitous in solar phenomena, remains however the celebrated 11 years Schwabe solar cycle. The working of the underlying clock is still unknown. R. Wolf already in 1859 [1] was the first to bring-up the possible planetary origin of the 11 years periodic behaviour of the sunspots, because of Jupiter's close orbital period (~11.8 yr). In fact, various investigations could establish a clear correlation between the Sun's cyclic dynamical behaviour and the planetary orbiting periods [2]. As the most obvious and promising potential mechanism behind such a planetary impact on the Sun, it has been considered gravitational tidal forces acting on the Sun, mainly by Jupiter; their periodicity 'drifts' towards the solar cycle, if a few other inner solar planets are also included when summing up their periodic tidal impact. However, it was realized that the estimated planetary tidal impact was extremely small to cause any significant change of the dynamic Sun [2], or even less to justify the origin of the enigmatic 11 years cycle. For this reason the planetary - Sun connection has been ignored for long time, while such a claim was also seen not only within astronomy, but rather instead within the frame of astrology [3]! Though, the significance of the correlation was high, since planetary tides follow a temporal pattern with a conspicuous correlation with the solar activity cycle, and therefore this challenging observation was not set *ad acta* [4,5].

   Hence, the many faces of the 11 years solar cycle may hold many important clues as to how the solar clock is working. To constrain the underlying mechanism(s), we discuss here an alternative scenario, which couples the dynamic Sun with the planets via the already introduced streams of dark constituents [6]. Further, dark disk configurations which co-rotate with the galaxy may contribute to the local dark matter flows [7]. Actually, this seems to be the only procedure left-over, in which the precise planetary periods can enter into the suggested 11 years scenario, explaining thus how the apparent 'communication' between the planets and the Sun is settled.
   .

2. **The new mechanism**

   In this work we suggest a new physical mechanism aiming to explain the solar cycle, which is cosmic in origin. It is based not on planetary torque, but on the gravitational lensing effect by the planets as they revolve with a constant orbital period around the Sun, entailing all the striking planetary periodic changes. In fact, they can focus gravitationally at the Sun's position slow moving incident dark matter (or any other exotic) constituents [6]. The flux enhancement and its duration depend on the relative alignment between the Sun, the planet(s) and the otherwise as yet invisible cosmic irradiation [8]. For a flux enhancement to occur, the incident irradiation of the solar system by any kind of feebly interacting particles must not be isotropic, arriving preferentially along the ecliptic plane. In this way, the planets may still leave *somehow* their imprints as the Sun's 11 years enigmatic activity rhythm. The bulk of the celebrated dark matter halo in our neighbourhood is not further considered here, since its origin goes back to the early Universe, and therefore it is isotropic. By contrast, for example, non-relativistic particles from point-like sources along / near the ecliptic plane like the celebrated "constellations", or, incident slow moving streams of dark matter or the like, can be gravitationally lensed towards the Sun by one or more planet(s), when a stream is properly co-aligned with the Sun and the planet(s). This can happen, because of the $v^{-2}$ - dependence of the lensing (=deflection) angle [8]. For example, the planets Jupiter and Earth can focus at the Sun's position incoming particles with speeds $v \leq 10^{-2}c$ and $v \leq 3 \cdot 10^{-3}c$, respectively, provided such particles propagate near the ecliptic, since most planets move coplanar (within a few degrees). We recall that relativistic particles ($v \approx c$) have focal lengths substantially larger than the orbital radius of Jupiter, even if the Sun is taken as the gravitational lens. But, over the last ~150 years, Jupiter's 11.8 years orbital revolution around the Sun was considered as the possible cause of the strikingly close ~11 years solar cycle, despite the rejected tidal mechanism (see e.g. [2]).

We note that the mentioned speeds of non-relativistic particles, resemble that of dark matter, but also dark constituents produced possibly in stars. Therefore, their direction of propagation can be influenced noticeably by the planetary gravitational fields. We mention, as a generic example, massive solar axions of the Kaluza-Klein (KK) type [10], which escape from the Sun with a mean velocity of about 0.6c. For the purpose of this work, it is reasonable to assume that a percentage of the solar KK axions (or the like) of about 1‰ leave the Sun with speeds below about 0.01c (see Figure 5 in ref. [10]). However, the fraction of such or other slowly moving exotica is not negligible, and it might be much more from other stars like pulsars, because of the much stronger gravity ($v_{escape} \approx 0.3c$).

It is of particular interest, the actual flux enhancement which can be expected by planetary gravitational focusing. Thus, it was shown recently [9] that Jupiter can cause, in the ideal case, a flux increase at its focal plane of as much as by a factor of ~$10^6$, assuming incident streaming particle candidates from the dark sector with the aforementioned speeds. As it was pointed out for KK- axions, it is not unreasonable to assume that such or other slow moving dark fluxes do exist; they may reach the solar system either as some sort of streaming dark matter, or, they may come from some point-like sources in the sky. Then, they can get (temporally) focused, by one or more planets, towards and interfere with the Sun. Apparently, similar gravitational lensing can take place between the planets, and other celestial bodies.

Thus, the sporadic planetary co-alignment repeats in precisely predictable time intervals. Coincidentally, the various planetary configurations elaborated for the tidal scenario can be taken over for this work, as it also gives the time-variable influx of focused directional dark constituents. Of course, this additional influx must interact with the Sun and cause a considerable influence, whatever the underlying process is at the end (see below).

***Some numerics*:** The possible existence of dark matter particle streams in the galactic halo has been already considered [6,7]. The Sagittarius Dwarf Elliptical Galaxy is a well studied case. The expected stream density at the Sun's position is a few % of the local dark halo, with stream velocities around $10^{-3}c$ [6,7]. Here we assume that streaming dark constituents make about 1% of the local ~0.3 GeV/cm$^3$ relic dark matter. With velocities around $10^{-2}$-$10^{-3}c$, the integrated energy flux reaching the Sun can be as much as $10^{30\pm1}$erg/s, if (temporarily / periodically) a ~$10^6$ times flux enhancement due to planetary gravitational focusing takes place. Such an external energy influx (up to ~$10^{-2}L_\odot$) is possibly not negligible. We mention, for comparison reasons, that the much less radiant energy emitted in X-rays by the solar corona (~$10^{24\pm2}$erg/s) cannot be overlooked, while known physics failed to explain its origin since several decades [12]. Keeping in mind the behaviour of axion(-like) particles [12,13], the energy deposit by a directional external dark irradiation of the Sun may take place spatiotemporally only at certain solar magnetized layers of specific density, etc. Though, the magnetic field is for particles like paraphotons redundant due to the kinetic mixing of the photon-to-paraphoton oscillation [13]. In addition, incident dark matter particles may be gravitationally captured and accumulated with time inside the Sun. Such or other processes may bring the Sun out of equilibrium short and/or long term, giving rise to the otherwise puzzling and unpredictable (local / global) solar activity. Some of the diverse exotica from the dark sector, like axions, paraphotons, WISPs, WIMPs, etc., may interact 'preferably' with the Sun, since its huge dynamic range of properties none Earth-bound detector can actually mimic. For example, if the additional energy deposit goes, e.g., via the Primakoff – effect [12,13], a fine-tuned spatiotemporal resonance between the rest mass of the dark constituents, the local solar plasma frequency (=energy), and/or eventually the local solar magnetic field, may occur *somewhere* inside the Sun or its atmosphere. Such an interaction occurring only with the Sun, it does not necessarily contradict Earth bound dark matter experiments. After all, they could not unravel a signature, because they have failed to pin down those necessary conditions within their very limited detector parameter values. Finally, in order to recuperate an 11 years solar cycle from the since ever suggestive 11.8 years Jupiter's orbital period, it suffices to consider the combined gravitational lensing effect with few more inner ones. Figure 1 shows actually an outstanding agreement between Schwabe's solar cycle periods and the planetary tides [11,2], which are used here as proxy for gravitational lensing.

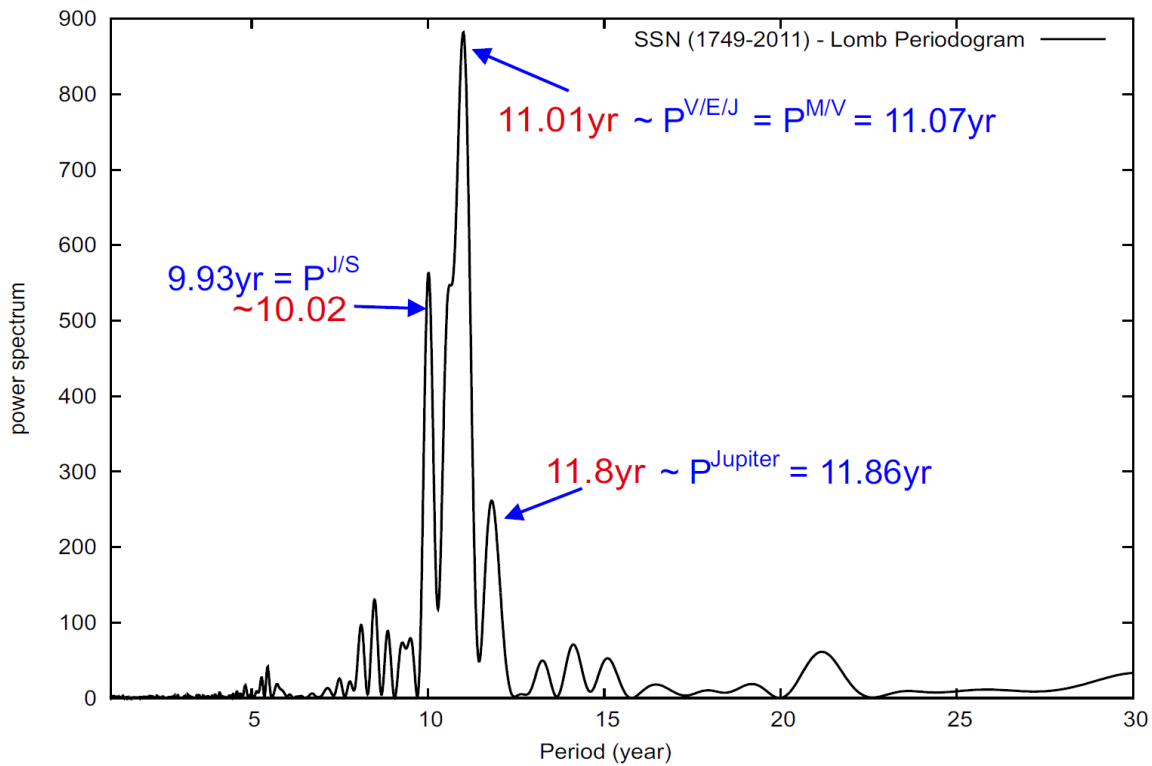

**Figure 1** The monthly average sunspot number reveals the existence of three peaks around 11 years (red) [11], which are all associated to planetary tides (blue). Tidal periods (**P**) of single and combined planets by Jupiter, Saturn, Venus, Earth, Mercury fit to planetary frequencies around 11 years. Note, in this work tidal timing is a proxy for gravitational lensing by the same planet(s). Courtesy, Nicola Scafetta (2013).

## 3. Discussion

The suggested planetary gravitational lensing scenario fits the characteristic timing of the as yet enigmatic solar cycle, which follows impressively the combined orbital rhythm of the inner planets; the Jupiter's period (11.8 years) is the most strikingly one close to the 11 years. The earlier suggestion, based on the tidal forces acting on the Sun by the various planetary configurations, failed to explain any reasonable impact on the Sun's workings. But, interestingly, most derived findings there, e.g., the planetary alignment(s) and period(s) of appearance, can be borrowed actually unmodified to corroborate the alternative scenario based on gravitational focusing of streaming constituents from the dark sector. The plethora of candidates like slow moving massive exotica, from axions and axion-like particles [12-15] to D-particles defects [16], which have been already discussed, are inspiring and may provide the energy input for the present solution of the 11 years solar clock.

Then, Wolf's suggestion was advanced for his time, since both, gravitational lensing and dark matter, were unknown. In fact, Jupiter provides the main driving force of the solar cycle, while the synergy with the other planets shifts slightly the 11.8 years period to the 11 years solar master clock. Actually, there is nothing else one could imagine beyond the assumed flows of dark particles, which may settle such an oscillatory behaviour for the Sun being identical with the combined planetary orbital rhythm. Then, it is not unreasonable to assume that the main as yet unidentified piece of the whole puzzle, i.e., some kind of dark streams do exist, showering the Sun, thanks to the intervening planets, periodically and probably also irregularly, in large quantities. Moreover, the mystery of the 11 years solar cycle might be pointing at the properties of the assumed dark streams towards the Sun near the ecliptic, whose intensity reaching the Sun gets occasionally enhanced by the planetary gravitational lenses. The same holds also for Earth-bound or some experiments in space, since they may profit from introducing in the data analysis the aforementioned period(s), or, the predicted time intervals with increased signal-to-noise ratio due to flux enhancement. This might allow to unravel an otherwise hidden signature.

## 4. Conclusions

It is suggested, that the mysterious 11 years solar cycle could be explained by incident particle streams (from the widely discussed dark sector) towards the Sun. More specifically, the flux of expected streaming dark matter component(s) beyond the isotropic local dark matter halo (~0.3 GeV/cm$^3$) with velocities around $10^{-3}$-$10^{-2}$c can be temporarily increased by the gravitational lensing potential of a single or more planets. For a constant influx of dark particles near the ecliptic, the combined planetary focusing efficiency shows surprisingly a periodicity of 11 years. In addition, dark streams varying with time could explain the fluctuations of the 11 years period in length and in amplitude. The same might hold for the unpredictable nature of puzzling, irregularly occurring solar events, which could also be another manifestation of the suggested scheme. Moreover, the expected external energy input to the Sun due to periodic flux enhancement is not negligible. This is true when a comparison is made with the total solar luminosity, but it is more suggestive, if such an additional external irradiation is compared with the several orders of magnitude weaker corona energy emission in X-rays, which is unexpected for a cool star like our Sun.

Interestingly, the coronal emission in X-rays shows also an 11 years cycle, though with a change in intensity by a factor of about $10^2$. This is to be compared with the corresponding amplitude variation of the bulk of the solar luminosity, which is only at the $10^{-3}$ level! The impressive 11 years coronal modulation must be seen on top of the already enigmatic origin of the coronal heating mechanism. Obviously, the 75 years old corona riddle becomes even more intriguing, as one has to explain, not only the puzzling temperature inversion occurring close to the photosphere, but also why the Sun emits in X-rays at all (and even more so above quiet magnetized regions), and, why its X-ray brightness changes with time following the mysterious 11 years clock. These mysteries may be interrelated, and they may or may not be of common origin. Therefore, in dark matter research the anyhow experimentally challenging detection in the (sub-)keV energy range, seems even more promising to pursue, as it might become *the* window to the (multifaceted?) dark sector. The energy overlap with the mysterious X-ray luminous Sun is certainly motivating.

*In summary,* the efficient planetary gravitational lensing of slow dark streams towards the Sun is suggested as the underlying mechanism, which drives the 11 years cycle. Actually, one may ask, since tidal effects have been excluded from further consideration, what else could fit to the striking 11 years solar rhythm reminiscent of the identical combined orbital period of Sun's inner planets? Following this scenario, the 11 years solar cycle with its many faces is the overlooked manifestation of streaming constituents from the dark sector. Hence, dark matter exotica show up not only on cosmic scales due to their prevailing gravitational force, but also on sizes like the solar system, or even much smaller. Then, there exists a preferred spatial direction in our neighbourhood, which is given by the flow of new exotica. Moreover, if (directional) dark matter constituents cause the mysterious and multifaceted 11 years cycle, this could provide the tool to design accordingly future direct dark matter searches, while aiming to unravel overlooked signatures by re-evaluating previous experiments / observations.